\documentclass[letterpaper, 10 pt, journal, twoside]{IEEEtran}

\IEEEoverridecommandlockouts

\usepackage{graphics} 
\usepackage{epsfig} 
\usepackage{mathptmx} 
\usepackage{times} 
\usepackage{amsmath} 
\usepackage{amssymb}  
\usepackage{calrsfs}
\usepackage{booktabs}
\usepackage{algorithm, algorithmicx}

\DeclareMathAlphabet{\pazocal}{OMS}{zplm}{m}{n}
\newcommand{\norm}[1]{\left\lVert#1\right\rVert}
\newtheorem{thm}{Theorem}
\newtheorem{lm}{Lemma}
\newtheorem{rmk}{Remark}

\title{
Subspace Identification of Linear Time-Periodic Systems with Periodic Inputs
}

\author{Mingzhou Yin, Andrea Iannelli, Roy S. Smith
\thanks{This work is supported by the Swiss National Science Foundation under grant no.: 200021\_178890.}
\thanks{Mingzhou Yin, Andrea Iannelli, and Roy S. Smith are with the Automatic Control Laboratory, Swiss Federal Institute of Technology (ETH Zurich), Physikstrasse 3, 8092 Zurich, Switzerland, \texttt{\{myin,iannelli,rsmith\} @control.ee.ethz.ch.}}
\thanks{\textcopyright\ 2020 IEEE. Personal use of this material is permitted. Permission from IEEE must be obtained for all other uses, in any current or future media, including reprinting/republishing this material for advertising or promotional purposes, creating new collective works, for resale or redistribution to servers or lists, or reuse of any copyrighted component of this work in other works.}
}

\begin{document}

\maketitle
\thispagestyle{empty}
\pagestyle{empty}

\begin{abstract}
This paper proposes a new methodology for subspace identification of linear time-periodic (LTP) systems with periodic inputs. This method overcomes the issues related to the computation of frequency response of LTP systems by utilizing the frequency response of the time-lifted system with linear time-invariant structure instead. The response is estimated with an ensemble of input-output data with periodic inputs. This allows the frequency-domain subspace identification technique to be extended to LTP systems. The time-aliased periodic impulse response can then be estimated and the order-revealing decomposition of the block-Hankel matrix is formulated. The consistency of the proposed method is proved under mild noise assumptions. Numerical simulation shows that the proposed method performs better than multiple widely-used time-domain subspace identification methods when an ensemble of periodic data is available.
\end{abstract}

\begin{IEEEkeywords}
Subspace methods, identification, time-varying systems.
\end{IEEEkeywords}

\section{Introduction}

\IEEEPARstart{L}{inear} time-periodic (LTP) systems are systems with periodically varying linear dynamics. Periodicity is observed in various applications, e.g. \cite{Wood_2018,Khosravi_2017,2000}. More importantly, LTP systems serve as an intermediate step to capture more general representations than linear time-invariant (LTI) systems, for example, linear parameter-varying (LPV) systems \cite{Felici_2007,Goos_2014,cox2018towards} and nonlinear systems along limit cycles \cite{Allen_2009}.

This paper focuses on identifying state-space LTP models from input-output data. Based on early work from \cite{wereley1990analysis}, methods were developed to estimate nonparametric models of the harmonic transfer function \cite{Allen_2009,Shin_2005}. This input-output model can then be realized as state-space form \cite{Goos_2014}. The most successful state-space method is probably the time-domain subspace identification method \cite{Verhaegen_1995}, which extends naturally from its LTI counterpart \cite{Van_Overschee_1996}. This method, along with a similar version in \cite{Hench_1995}, has contributed to a number of successful applications (e.g., \cite{Felici_2007,Sefidmazgi_2016}), especially in identifying LPV systems where modern subspace techniques have been incorporated \cite{cox2018towards}. On the other hand, the frequency-domain subspace formulation for LTP systems has not been investigated until the recent paper \cite{Uyanik_2019} based on frequency lifting. However, the method is limited to single-input and single-output (SISO) systems with multi-sinusoidal inputs. In addition, the frequency grid needs to be specially designed to avoid overlaps between different periodic harmonics. This work aims to propose an alternative framework that is compatible with more general inputs and systems.

The importance of developing frequency domain methods in system identification lies in the advantage of using periodic inputs in identification experiments. As discussed in \cite{Schoukens_1994}, periodic input design has a number of advantages compared to random input design, including avoiding initial state estimation and easier time-domain averaging. However, the frequency response behaviour of LTP systems differs significantly from that of LTI systems \cite{wereley1990analysis}. Most prominently, the independence of the frequency response at different frequencies, a property that is fundamental to frequency-domain identification of LTI systems, does not hold for LTP systems. This prevents straightforwardly applying LTI techniques to frequency-domain identification of LTP systems. The key idea of this work is to use the frequency response of time-lifted systems with LTI structure to overcome this limitation.

The technical contribution of the paper is the proposal of a novel frequency-domain subspace identification method for multi-input multi-output (MIMO) LTP systems. First, the frequency response of the lifted system is identified by the generalized empirical transfer function estimate (ETFE). Then, the method extends the frequency-domain subspace identification method in \cite{McKelvey_1996} to LTP systems. By utilizing the frequency response of the lifted system, the time-aliased periodic impulse response of the original LTP system can be obtained by a linear mapping. The time-aliased periodic impulse response then leads to an order-revealing decomposition of LTP systems with block-Hankel structure. This is followed by a conventional subspace routine that identifies the range space of the extended observability matrix by performing a singular value decomposition. This algorithm is proven to be consistent under a general class of output noise. Compared to \cite{Uyanik_2019}, the main advantages are that it can be applied to MIMO systems and that generic periodic inputs can be used. However, compared with previous time-domain methods which use arbitrary input-output data sequence(s), this method requires an ensemble of periodic identification data that are harmonic with the fundamental frequency of the system. Finally, the proposed algorithm is compared to the time-domain method by numerical simulation to show its advantage with periodic identification data. The consistency property is also verified in simulation.


\section{Problem Statement}
\label{sec:form}

Consider a discrete-time strictly-causal LTP system with the following minimal state-space model
\begin{equation}
\begin{cases}
x(t+1)&=\ A_t x(t)+B_t u(t)\\
\hfil y(t)&=\ C_t x(t)
\label{eq:sys}
\end{cases},
\end{equation}
where $x \in \mathbb{R}^{n_x}$, $u \in \mathbb{R}^{n_u}$, and $y \in \mathbb{R}^{n_y}$ are the states, inputs, and outputs respectively. The time-varying matrices $A_t=A_{t+P}$, $B_t=B_{t+P}$, $C_t=C_{t+P}$ are periodic state-space matrices of appropriate dimensions, and $P$ is the period length. Denote the collection of unique $A$-matrices as $A=[A_0^\top\ A_1^\top\ \cdots\ A_{P-1}^\top]^\top$, similarly for $B$ and $C$. The monodromy matrix of the system is defined as $\Psi_{A,t}=A_{t-1}A_{t-2}\cdots A_{t-P}$ \cite{bittanti2009periodic}.
The periodic impulse response of the system is defined as
\begin{equation}
g^t_{r}=C_t A_{t-1} A_{t-2} \cdots A_{t-r+1} B_{t-r}\in \mathbb{R}^{n_y\times n_u},
\label{eq:imp}
\end{equation}
where $t$ is the tag time of the impulse response and $r> 0$ is the input-output lag. The response $g^t_{r}$ is $P$-periodic with respect to $t$. 

In the remainder of the paper, the following system identification problem is considered:

	\textbf{Given:} $J$ input-output data sequences of system (\ref{eq:sys}) with periodic inputs of length $NP$, where $J\geq Pn_u$. The inputs and noise contaminated outputs are denoted as $u^i(t)$ and $y^i(t)=y_0^i(t)+w^i(t)$ respectively, where $t = 0,1,\cdots,NP-1$ denotes the measured time instants, $i=1,2,\cdots,J$ denotes the index of the experiments, $y_0^i(t)$ is the noise-free output, and $w^i(t)$ is the unknown noise.
	
	\textbf{Assumptions:} 1) the system is stable, i.e., $\rho(\Psi_{A,t})<1$; 2) the noise is i.i.d. across experiments, and not correlated with the inputs, i.e., $\mathbb{E}\left[(u(t_1)-\mathbb{E}\left[u(t_1)\right])w^\top(t_2)\right]=0$; 3) the noise is zero mean with fast-decaying covariances $\sum_{\tau=1}^\infty\left|\tau\cdot\mathbb{E}\left[w_p(t)w_p(t-\tau)\right]\right|=c_p<\infty$, where $w^i_p$ is the $p$-th element of $w^i$; 4) the period length $P$ is known. 
	
	\textbf{Objective:} estimate a state-space LTP model that is equivalent to (\ref{eq:sys}) up to a similarity transform.

\section{Frequency Response of LTP Systems}
\label{sec:freq}

An important characteristic of LTP systems is that, unlike LTI systems, an input with spectral content at frequency $\omega$ will generate an output response not only at $\omega$, but also at a series of other harmonics $\omega+2k\pi/P,\ k\in\mathbb{Z}$ \cite{wereley1990analysis}. Thus, the frequency response at a particular frequency $\omega$ is not a complex gain, but a function $G_\omega(\omega+2k\pi/P)$ of $k$. This function-valued frequency response can be estimated at individual frequencies with a technique known as frequency lifting \cite{Uyanik_2019}. However, this method is very restrictive in input design, in that only carefully designed multi-sinusoidal inputs can be applied to ensure no overlap of harmonics with different input frequency content. In this paper, a time-lifted method is considered for arbitrary periodic inputs of length $NP$, $N\in\mathbb{N}_+$. As will be seen in Section~\ref{sec:ltp}, this method is useful in extending the available frequency-domain subspace identification algorithm to LTP systems. For the rest of the paper, the term lifting refers to time-lifting.

Lifting is one of the most common LTI reformulations of LTP systems. In the lifted system, the inputs and outputs of one whole period in the LTP system are concatenated as the new inputs and outputs
\begin{equation}
\tilde{u}^i(k) =
\begin{bmatrix}
{u^i}^\top\!\!(kP)&{u^i}^\top\!\!(kP+1)&\cdots&{u^i}^\top\!\!(kP+P-1)
\end{bmatrix}^\top,
\label{eqn:lift}
\end{equation}
similarly for $\tilde{y}^i(k)$ and $\tilde{w}^i(k)$. The result is a structured LTI system of $P$-times larger input and output dimensions and $P$-times slower. The state dimension remains the same.

In this way, the frequency response matrix of the lifted LTI system $G(\text{e}^{j\omega_k})$ can be used as the frequency response data of the original LTP system. It is shown in Section~4.3 of \cite{Bittanti_2000} that the frequency response of the lifted system is given by
\begin{equation}
G_{l,m}(\text{e}^{j\omega_k})=\sum_{s=0}^{\infty}g^l_{sP+l-m}\exp{\left(-j\omega_k s\right)},
\end{equation}
where $G_{l,m}\in\mathbb{C}^{n_y\times n_u}$ denotes the $l$-$m$th block element of $G$, $\omega_k=\frac{2\pi k}{N}$, $k=0,1,\cdots,N-1$. Note that, due to the strict causality assumption of (\ref{eq:sys}), $g^t_r=0$ for all non-strictly-causal impulse response coefficients, that is for $r\leq 0$.

Despite its LTI structure, frequency response estimation of the lifted MIMO system is not a trivial problem as conventional methods such as swept-sine and multi-sines \cite{Dobrowiecki_2006} are not applicable to lifted LTP systems as the input channels cannot be excited separately, since they come from the same input sequence. Therefore, we propose the following generalized ETFE $\hat{G}(\text{e}^{j\omega_k})$ similar to \cite{Dobrowiecki_2006} but from an ensemble of time-domain identification data with periodic inputs.

We apply the discrete Fourier transform (DFT) on each channel of the lifted inputs and outputs,
\begin{equation}
    U_i(\text{e}^{j\omega_k})=\sum_{n=0}^{N-1}\tilde{u}^i(n) \exp\left(-j\frac{2\pi n k}{N}\right),
    \label{eqn:etfe1}
\end{equation}
and similarly for $Y_i(\text{e}^{j\omega_k})$ and $W_i(\text{e}^{j\omega_k})$. Then the frequency response estimate is given as
\begin{equation}
    \hat{G}(\text{e}^{j\omega_k})=\tilde{Y}(\text{e}^{j\omega_k})\tilde{U}^\dagger(\text{e}^{j\omega_k}),
    \label{eqn:etfe2}
\end{equation}
where
\begin{equation}
    \tilde{U}(\text{e}^{j\omega_k})=
    \begin{bmatrix}
    U_1(\text{e}^{j\omega_k})&U_2(\text{e}^{j\omega_k})&\cdots&U_J(\text{e}^{j\omega_k})
    \end{bmatrix},
\end{equation}
similarly for $\tilde{Y}(\text{e}^{j\omega_k})$ and $\tilde{W}(\text{e}^{j\omega_k})$. Here, for the right pseudo-inverse to be well defined, $\tilde{U}(\text{e}^{j\omega_k})$ needs to have full row rank, which requires $J\geq Pn_u$.

The estimate (\ref{eqn:etfe2}) generalizes the ETFE for the SISO case
\begin{equation}
    \hat{G}(\text{e}^{j\omega_k})=\frac{Y(\text{e}^{j\omega_k})}{U(\text{e}^{j\omega_k})}
\end{equation}
with multiple experiments to satisfy the persistency of excitation requirement for MIMO systems. We will show that this estimate has similar properties to the ETFE, i.e., it is unbiased with bounded covariances and the estimation errors are independent across different frequencies. Note that for notational simplicity, a MISO structure is considered in the proof but the same properties hold for the MIMO system with covariance of the vectorized $\hat{G}(\text{e}^{j\omega_k})$.

\begin{lm}
     Given the assumptions in Section~\ref{sec:form}, the frequency response estimate (\ref{eqn:etfe2}) has the following properties:
     \begin{enumerate}
         \item 
        $\mathbb{E}\left[\hat{G}(\text{e}^{j\omega_k})\right]=G(\text{e}^{j\omega_k})$,
        \item $\text{Cov}\left[\hat{G}^{(p)}\right]=\left(\Phi_{w_p}+\rho_p(N)\right)\left(\tilde{U}^\dagger\right)^\mathsf{H}\tilde{U}^\dagger$, where $\hat{G}^{(p)}$ denotes the $p$-th row of $\hat{G}$, $\Phi_{w_p}$ is the power spectral density of the $p$-th element of $w$, and $|\rho_p(N)|\leq 2c_p/N$. Note that the frequency dependence is omitted for simplicity.
        \item estimates at different frequencies are independent.
     \end{enumerate}
     \label{lm:1}
\end{lm}

\begin{IEEEproof}
    Decompose the lifted MIMO system into $P n_y$ multiple-input single-output (MISO) systems with
    \begin{equation}
        G(\text{e}^{j\omega_k})=\begin{bmatrix}
        {G^{(1)}}^{\!\top}\!\!(\text{e}^{j\omega_k})&{G^{(2)}}^{\!\top}\!\!(\text{e}^{j\omega_k})&\cdots&{G^{(Pn_y)}}^{\!\top}\!\!(\text{e}^{j\omega_k})
        \end{bmatrix}^\top,
    \end{equation}
    and similarly for $\hat{G}(\text{e}^{j\omega_k})$. Then,
    \begin{equation}
        \tilde{Y}^{(p)}(\text{e}^{j\omega_k})=G^{(p)}(\text{e}^{j\omega_k})\tilde{U}(\text{e}^{j\omega_k})+\tilde{W}^{(p)}(\text{e}^{j\omega_k}),
    \end{equation}
    \begin{equation}
        \hat{G}^{(p)}(\text{e}^{j\omega_k})=G^{(p)}(\text{e}^{j\omega_k})+\tilde{W}^{(p)}(\text{e}^{j\omega_k})\tilde{U}^\dagger(\text{e}^{j\omega_k}),
    \end{equation}
    where $\tilde{Y}^{(p)}(\text{e}^{j\omega_k})$, $\tilde{W}^{(p)}(\text{e}^{j\omega_k})$ denote the $p$-th row of $\tilde{Y}(\text{e}^{j\omega_k})$, $\tilde{W}(\text{e}^{j\omega_k})$ respectively.
    With zero-mean noise, the estimate is unbiased
    \begin{equation}
    \begin{aligned}
        \mathbb{E}\left[\hat{G}^{(p)}(\text{e}^{j\omega_k})\right]&=G^{(p)}(\text{e}^{j\omega_k})+\mathbb{E}\left[\tilde{W}^{(p)}(\text{e}^{j\omega_k})\right]\tilde{U}^\dagger(\text{e}^{j\omega_k})\\&=G^{(p)}(\text{e}^{j\omega_k}).
    \end{aligned}
    \end{equation}
    The covariance of the estimate is given by
    \begin{equation}
    \begin{aligned}
        \mathbb{E}\left[\left(\hat{G}^{(p)}(\text{e}^{j\omega_k})-G^{(p)}(\text{e}^{j\omega_k})\right)^\mathsf{H}\left(\hat{G}^{(p)}(\text{e}^{j\omega_m})-G^{(p)}(\text{e}^{j\omega_m})\right)\right]\\
        =\left(\tilde{U}^\dagger(\text{e}^{j\omega_k})\right)^\mathsf{H}\mathbb{E}\left[\left(\tilde{W}^{(p)}(\text{e}^{j\omega_k})\right)^\mathsf{H}\tilde{W}^{(p)}(\text{e}^{j\omega_m})\right]\tilde{U}^\dagger(\text{e}^{j\omega_m}).
    \end{aligned}
        \label{eqn:etfep1}
    \end{equation}
    From Section~6.3 of \cite{LjungBook2} and the independence across different experiments, we have
    \begin{multline}
        \mathbb{E}\left[\left(\tilde{W}^{(p)}(\text{e}^{j\omega_k})\right)^\mathsf{H}\tilde{W}^{(p)}(\text{e}^{j\omega_m})\right]\\=\begin{cases}
        \left(\Phi_{w_p}(\text{e}^{j\omega_k})+\rho_p(N)\right)I,&\omega_k=\omega_m\\
        \mathbf{0},&\omega_k\neq\omega_m
        \end{cases},
        \label{eqn:etfep2}
    \end{multline}
    Substituting (\ref{eqn:etfep2}) into (\ref{eqn:etfep1}) completes the proof.
\end{IEEEproof}

\begin{rmk}
When $P$ is unknown, cross-validation can be performed by obtaining the generalized ETFE estimate with lifting structures of different $P$.
\end{rmk}

\section{Frequency-Domain Subspace Identification of LTP Systems}
\label{sec:decom}

To develop the frequency-domain subspace identification method for LTP systems based on the frequency response of the lifted system, we first examine the algorithm for that of LTI systems. This is briefly summarized based on the uniformly spaced data case in \cite{McKelvey_1996}.

\subsection{The algorithm for LTI Systems}

Suppose $M$ frequency response data $G_k$ are given on uniformly spaced frequencies $\omega_k=2\pi k/M,k=0,1,\cdots,M-1$. First, apply the inverse discrete Fourier transform (IDFT) on $G_k$,
\begin{equation}
h_r = \frac{1}{M}\sum_{k=0}^{M-1}G_k\cdot \exp\left(j\frac{2\pi r k}{M}\right),\ r=1,2,\cdots,M.
\end{equation}
The sequence $h_t$ is then the time-aliased impulse response of the system,
\begin{equation}
h_r = \sum_{i=0}^{\infty}g_{r+iM}.
\end{equation}
Based on this result, the block-Hankel matrix of $h_t$ has the following decomposition that reveals the order of the system.
\begin{equation}
\begin{aligned}
H &= 
\begin{bmatrix}
h_1&h_2&\cdots&h_r\\
h_2&h_3&\cdots&h_{r+1}\\
\vdots&\vdots&\ddots&\vdots\\
h_q&h_{q+1}&\cdots&h_{r+q-1}\\
\end{bmatrix}\\
&=
\begin{bmatrix}
C\\CA\\\vdots\\CA^{q-1}
\end{bmatrix}
(I-A^M)^{-1}
\begin{bmatrix}
B&AB&\cdots&A^{r-1}B
\end{bmatrix}
,
\end{aligned}
\label{eqn:decom}
\end{equation}
Thus, the extended observability matrix of the system can be identified up to a similarity transform from the range space of $H$ by singular value decomposition and truncation. The order of the estimated system can be determined by thresholding or cross-validation.

\subsection{Order-revealing decomposition for LTP systems}
\label{sec:ltp}

With the frequency response of the lifted system, the order-revealing decomposition analogous to (\ref{eqn:decom}) can be developed for LTP systems.

Take the IDFT of $\hat{G}_{l,m}(\text{e}^{j\omega_k})$ in (\ref{eqn:etfe2}),
\begin{equation}
\begin{aligned}
    w_{l,m}(n) &= \frac{1}{N}\sum_{k=0}^{N-1}\hat{G}_{l,m}(\text{e}^{j\omega_k}) \exp\left(j\frac{2\pi n k}{N}\right)\\
    &=\frac{1}{N}\sum_{k=0}^{N-1}\sum_{s=0}^{\infty}g^l_{sP+l-m}\exp{\left(-j\frac{2\pi (s-n) k}{N}\right)}.
\end{aligned}
\label{eqn:ifdt}
\end{equation}
Since the summation over $k$ is on the whole unit circle, it is only non-zero when $s-n=iN,\ i\in \mathbb{N}$. We have
\begin{equation}
    w_{l,m}(n) = 
    \begin{cases}
        \sum_{i=0}^\infty g^l_{(iN+n)P+l-m},&nP+l-m > 0,\\
        \sum_{i=0}^\infty g^l_{(iN+N+n)P+l-m},&nP+l-m \leq 0,
    \end{cases}.
    \label{eqn:h1}
\end{equation}
Define the time-aliased periodic impulse response as
\begin{equation}
    h^t_{r} = \sum_{i=0}^\infty g^t_{r+iNP},\ r=1,2,\cdots,NP.
    \label{eqn:h2}
\end{equation}
According to the definition of $g^t_{r}$ (\ref{eq:imp}), $\forall p=0,1,\cdots,r-1$,
\begin{equation}
    h^t_{r} = C_t A_{t-1} \cdots A_{t-p} \left(I-\Psi_{A,(t-p)}^{N}\right)^{-1}A_{t-p-1}\cdots A_{t-r+1} B_{t-r}.
\end{equation}
Therefore, the periodic block-Hankel matrix of $h^t_{r}$ can be decomposed as follows
\begin{equation}
\begin{aligned}
H_p^\tau &= 
\begin{bmatrix}
h_1^\tau&h_2^\tau&\cdots&h_{r}^\tau\\
h_2^{\tau+1}&h_3^{\tau+1}&\cdots&h_{r+1}^{\tau+1}\\
\vdots&\vdots&\ddots&\vdots\\
h_{q}^{\tau+P-1}&h_{q+1}^{\tau+P-1}&\cdots&h_{q+r-1}^{\tau+P-1}\\
\end{bmatrix}\\
&=\pazocal{O}^\tau_{q}\left(I-\Psi_{A,\tau}^{N}\right)^{-1}\pazocal{C}^\tau_{r}
,
\end{aligned}
\label{eqn:decomltp}
\end{equation}
where $q+r-1\leq NP$, and
\begin{equation}
\pazocal{C}^\tau_{s}=
\begin{bmatrix}
B_{\tau-1} &\!\! A_{\tau-1}B_{\tau-2} &\!\! \cdots &\!\! A_{\tau-1}\cdots A_{\tau-s+1}B_{\tau-s}
\end{bmatrix}\in \mathbb{R}^{n_x\times sn_u}
\end{equation}
\vspace*{0pt}
\begin{equation}
\pazocal{O}^\tau_{s}=
\begin{bmatrix}
C_{\tau} \\ C_{\tau+1}A_\tau \\ \vdots \\ C_{\tau+s-1}A_{\tau+s-2}\cdots A_\tau
\end{bmatrix}\in \mathbb{R}^{sn_y\times n_x},
\label{eq:O}
\end{equation}
are the extended controllability and observability matrices of LTP systems respectively \cite{bittanti2009periodic}.
By selecting $q,r$ such that $qn_y\geq n_x$, $rn_u\geq n_x$, together with the minimality and stability of the system, we have
\begin{equation}
\begin{aligned}
    \text{rank}\left(H_p^\tau\right) &= \text{rank}\left(\pazocal{O}^\tau_{q}\right) = \text{rank}\left(\left(I-\Psi_{A,\tau}^{N}\right)^{-1}\right) \\ &=\text{rank}\left(\pazocal{C}^\tau_{r}\right)=n_x.
\end{aligned}
\end{equation}
Note that the rank requirements on $q$ and $r$ put a lower bound on $N$. Then the range space of $H_p^\tau$ coincides with that of $\pazocal{O}^\tau_{q}$. Thus, $\pazocal{O}^\tau_{q}$ can be identified, up to a similarity transform, by performing singular value decomposition on $H_p^\tau$. From the extended observability matrix, the matrices $A$ and $C$ can be estimated by the same shifting method as in the time-domain subspace identification of LTP systems \cite{Verhaegen_1995}. The input matrix $B$ can be estimated by least-squares fit to the time-aliased impulse response.

\section{Algorithm \& Consistency Analysis}
\label{sec:algo}

Built on the decomposition (\ref{eqn:decomltp}), we propose Algorithm~\ref{al:1} for frequency-domain subspace identification of LTP systems with periodic inputs.

\begin{algorithm}[htb]
	\caption{Frequency-domain subspace identification of LTP systems with periodic inputs}
	\begin{algorithmic}[1]
		\State Lift the input-output data $u^i(t)$, $y^i(t)$ to $\tilde{u}^i(k)$, $\tilde{y}^i(k)$, $k=0,1,\cdots,N-1$, as in (\ref{eqn:lift}).
		\State Estimate the frequency response $\hat{G}(\text{e}^{j\omega_k})$ of the lifted system from $\tilde{u}(k)$, $\tilde{y}(k)$ by (\ref{eqn:etfe1}) and (\ref{eqn:etfe2}).
		\State Apply the IDFT on each block element $\hat{G}_{l,m}(\text{e}^{j\omega_k})$ of $\hat{G}(\text{e}^{j\omega_k})$ according to (\ref{eqn:ifdt}) and denote it as $\hat{w}_{l,m}(n)$.
		\State Construct the time-aliased periodic impulse response $\left\{\hat{h}_r^t\right\}$, $r=1,2,\cdots,NP$ by rearranging elements in $\hat{w}_{l,m}(n)$ according to (\ref{eqn:h1}) and (\ref{eqn:h2}).
		\State Construct $\hat{H}_p^\tau$ for $\tau=0,1,\cdots,P-1$, according to (\ref{eqn:decomltp}).
		\State Calculate the singular value decomposition of $\hat{H}_p^\tau$ for $\tau=0,1,\cdots,P-1$
		\begin{equation}
		    \hat{H}_p^\tau=\hat{U}_\tau\hat{\Sigma}_\tau\hat{V}_\tau^\top.
		\end{equation}
		\State Determine a system order $n_x$ and define $\hat{U}_\tau=\left[\hat{U}^s_\tau\ \hat{U}^o_\tau\right]$, where $\hat{U}^s_\tau\in \mathbb{R}^{qn_y \times n_x}$.
		\State The estimated state-space model is given as
		\begin{equation}
		\hat{A}_\tau = (J_1 \hat{U}^s_{\tau+1})^\dagger J_2\hat{U}^s_\tau,\ 
		\hat{C}_\tau = J_3\hat{U}^s_\tau, \tau=0,1,\cdots,P-1,
		\end{equation}
		\begin{equation}
		\hat{B} = \text{arg}\underset{B}{\text{min}}\ \sum_{r=1}^{NP}\sum_{\tau=0}^{P-1} \norm{\hat{h}_r^\tau-\hat{Q}_r^\tau B_{\tau-r}}_F^2,
		\label{eqn:ls}
		\end{equation}
		where $\hat{U}^s_P=\hat{U}^s_0$, $
		    J_1 =
		    \begin{bmatrix}
		        I_{(q-1)n_y}&\mathbf{0}_{(q-1)n_y\times n_y}
		    \end{bmatrix}$, $
		    J_2 =
		    \begin{bmatrix}
		        \mathbf{0}_{(q-1)n_y\times n_y}&I_{(q-1)n_y}
		    \end{bmatrix}$, $
		    J_3 =
		    \begin{bmatrix}
		        I_{n_y}&\mathbf{0}_{n_y\times (q-1)n_y}
		    \end{bmatrix}$, $
        \hat{Q}_r^\tau = \hat{C}_\tau\left(I-\Psi_{\hat{A},\tau}^{N}\right)^{-1} \hat{A}_{\tau-1} \cdots \hat{A}_{\tau-r+1}$.
	\end{algorithmic}
	\label{al:1}
\end{algorithm}

The computational complexity of Algorithm~\ref{al:1} is dominated by solving the least squares problem (\ref{eqn:ls}), which has a complexity of $O(n_x^2\cdot n_u^2\cdot N\cdot P^2)$.

We will show the following consistency property of Algorithm~\ref{al:1}.

\begin{thm}
Let $A_t$, $B_t$, and $C_t$ define the minimal LTP state-space model (\ref{eq:sys}). Let $\hat{A}_t$, $\hat{B}_t$, and $\hat{C}_t$ be the estimated state matrices by Algorithm~\ref{al:1}. Given the assumptions in Section~\ref{sec:form}, there exist nonsingular periodic matrices $T_t \in \mathbb{R}^{n_x\times n_x}$, $T_t=T_{t+P}$ such that w.p. 1,
\begin{equation}
    \lim_{N\to\infty}\norm{
    \begin{bmatrix}
    A_t&B_t\\
    C_t&\mathbf{0}
    \end{bmatrix}
    -
    \begin{bmatrix}
    T_{t+1}&\mathbf{0}\\
    \mathbf{0}&I
    \end{bmatrix}
    \begin{bmatrix}
    \hat{A}_t&\hat{B}_t\\
    \hat{C}_t&\mathbf{0}
    \end{bmatrix}
    \begin{bmatrix}
    T_t^{-1}&\mathbf{0}\\
    \mathbf{0}&I
    \end{bmatrix}
    }_F=0,
\end{equation}
for a fixed choice of $q,r$.
\label{thm:1}
\end{thm}

\begin{IEEEproof}
    Let $\Delta G(\text{e}^{j\omega_k})=\hat{G}(\text{e}^{j\omega_k})-G(\text{e}^{j\omega_k})$, $\Delta w_{l,m}(n)=\hat{w}_{l,m}(n)-w_{l,m}(n)$. We have
    \begin{equation}
        \Delta w_{l,m}(n) = \frac{1}{N}\sum_{k=0}^{N-1}\Delta G_{l,m}(\text{e}^{j\omega_k}) \exp\left(j\frac{2\pi n k}{N}\right),
    \end{equation}
    which can be seen as the sample mean of zero-mean independent random variables \cite{McKelvey_1996}. From Lemma~\ref{lm:1}, we know that the covariances of the random variables are bounded. Thus, according to the law of large numbers,
    \begin{equation}
	    \lim_{N\to\infty}\Delta w_{l,m}(n)=0,\ \text{w.p. }1,
    \end{equation}
    Then let $\Delta h_r^t = \hat{h}_r^t-h_r^t$, $\Delta H_p^\tau = \hat{H}_p^\tau-H_p^\tau$. We have
	\begin{equation}
	    \lim_{N\to\infty}\Delta h_r^t=0,\ \text{w.p. }1,
	\end{equation}
	which implies that, for $\tau=0,1,\cdots,P-1$,
	\begin{equation}
	    \lim_{N\to\infty}\norm{\Delta H_p^\tau}_F=0,\ \text{w.p. }1.
	    \label{eqn:c1}
	\end{equation}

	Let $\norm{\Delta H_p^\tau}_F\leq \epsilon$. According to the proof of Lemma 4 in \cite{McKelvey_1996}, there exist a matrix $P_\tau$ satisfying $\norm{P_\tau}_F\leq 4\epsilon/\sigma_{n_x}(H_p^\tau)$ and a non-singular matrix $T_\tau$ such that
	\begin{equation}
	    \hat{U}^s_\tau=(U^s_\tau+U^o_\tau P_\tau)T_\tau,
	\end{equation}
	where $H_p^\tau=\left[U^s_\tau\ U^o_\tau\right]\Sigma_\tau V_\tau^\top$. Then, we have
	\begin{equation}
	\begin{aligned}
		T_{\tau+1}\hat{A}_\tau T_\tau^{-1} &= \left(J_1 (U^s_{\tau+1}+U^o_{\tau+1} P_{\tau+1})\right)^\dagger J_2(U^s_\tau+U^o_\tau P_\tau),\\
		\hat{C}_\tau T_\tau^{-1} &= J_3(U^s_\tau+U^o_\tau P_\tau).
	\end{aligned}
	\end{equation}
	Note that
	\begin{equation}
	J_1 U^s_{\tau+1} A_\tau = J_2 U^s_\tau.\ 
	C_\tau = J_3 U^s_\tau,
	\end{equation}
	Then, from Theorem~5.3.1 in \cite{golub2012matrix} on the sensitivity of the least squares estimate, for a sufficiently small $\epsilon$ such that the regressor does not lose rank, there exists constants $c_\tau,c'_\tau$, such that
	\begin{equation}
	\begin{aligned}
		\norm{T_{\tau+1}\hat{A}_\tau T_\tau^{-1}-A_\tau}_F &\leq c_\tau \epsilon\\
		\norm{\hat{C}_\tau T_\tau^{-1}-C_\tau}_F &\leq c'_\tau \epsilon.
	    \label{eqn:c2}
	\end{aligned}
	\end{equation}
	For the estimate of $\left\{\hat{B}_\tau\right\}$ (\ref{eqn:ls}), let
    \begin{equation}
    Q_r^\tau = C_\tau\left(I-\Psi_{A,\tau}^{M}\right)^{-1} A_{\tau-1} \cdots A_{\tau-r+1}.
    \end{equation}
    Then a simple calculation shows that
    \begin{equation}
		\norm{\hat{Q}_r^\tau T_{\tau-r+1}^{-1}-Q_r^\tau}_F = O(\epsilon).
	\end{equation}
	Since $\Delta h_r^t=O(\epsilon)$, again from Theorem~5.3.1 in \cite{golub2012matrix}, for a sufficiently small $\epsilon$,
    \begin{equation}
		\norm{T_{\tau-r+1}\hat{B}_{\tau-r}-B_{\tau-r}}_F = O(\epsilon).
	\end{equation}
	The above equation, together with (\ref{eqn:c1}) and (\ref{eqn:c2}) completes the proof.
\end{IEEEproof}

\section{Numerical Examples}
\label{sec:ex}

In this section, the proposed algorithm is tested against multiple time-domain subspace identification algorithms for LTP systems with two numerical examples. Example~1 is based on the flapping dynamics of wind turbines, which is taken from \cite{Felici_2007}. The true dynamics of the system are given by
\begin{equation*}
\left[
\begin{array}{c|c}
A_0 & B_0 \\
\hline
C_0 & 0
\end{array}
\right]=
\left[
\begin{array}{cc|c}
0 & 0.0734 & -0.07221 \\
-6.5229 & -0.4997 & -9.6277 \\
\hline
1 & 0 & 0
\end{array}
\right],
\end{equation*}
\begin{equation*}
\left[
\begin{array}{c|c}
A_1 & B_1 \\
\hline
C_1 & 0
\end{array}
\right]=
\left[
\begin{array}{cc|c}
-0.0021 & 0 & 0 \\
-0.0138 & 0.5196 & 0 \\
\hline
0 & 0 & 0
\end{array}
\right],
\end{equation*}
where $n_x=2$, $n_y=n_u=1$, $P=2$. Example~2 is used in \cite{Hench_1995} with the dynamics
\begin{equation*}
\left[
\begin{array}{c|c}
A_0 & B_0 \\
\hline
C_0 & 0
\end{array}
\right]\!\!=\!\!
\left[
\begin{array}{cc|c}
1 & 1 & 0 \\
0 & 2 & 1 \\
\hline
1 & 0 & 0
\end{array}
\right],
\left[
\begin{array}{c|c}
A_1 & B_1 \\
\hline
C_1 & 0
\end{array}
\right]\!\!=\!\!
\left[
\begin{array}{cc|c}
\frac{1}{5} & 1 & 0 \\
0 & \frac{2}{5} & 1 \\
\hline
2 & 0 & 0
\end{array}
\right],
\end{equation*}
\begin{equation*}
\left[
\begin{array}{c|c}
A_2 & B_2 \\
\hline
C_2 & 0
\end{array}
\right]=
\left[
\begin{array}{cc|c}
3 & 1 & 1 \\
0 & 1 & 2 \\
\hline
1 & 1 & 0
\end{array}
\right],
\end{equation*}
where $n_x=2$, $n_y=n_u=1$, $P=3$. Both systems are then normalized to have an average steady-state gain of 1.

The compared algorithms are: 1) Algorithm~\ref{al:1} in this paper (\textit{Freq}), 2) the MOESP algorithm in \cite{Verhaegen_1995} (\textit{MOESP}), 3) the intersection algorithm in \cite{Hench_1995} (\textit{Int}), and 4) the CCA algorithm in Lemma~9.2 of \cite{cox2018towards} specialized for LTP systems (\textit{CCA}).

In both examples, the following simulation configuration and parameters are used. For each input-output data sequence, the systems are excited by periodic input of i.i.d. unit Gaussian entries $u(t)\sim \mathcal{N}(0,1)$ from zero initial conditions. The outputs are contaminated with i.i.d. unit Gaussian noise $y(t)=y_0(t)+w(t),w(t)\sim \mathcal{N}(0,1)$. The identification data are collected with $N=50$, $J=10\cdot P$ after the transient effect becomes negligible. The number of block-rows $q$ for the Hankel matrices in all methods are selected by cross validation. The system order $n_x$ is assumed to be known.

\begin{figure}[htbp]
\centering
\includegraphics[width=3.2in]{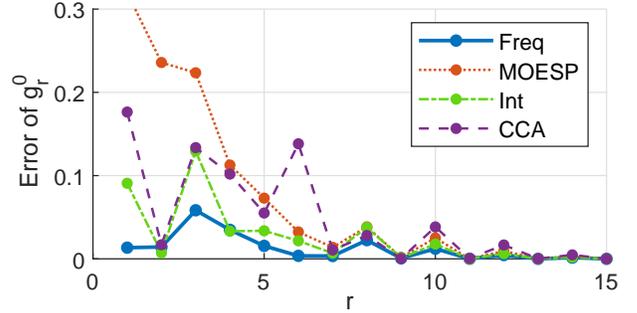}
\caption{Errors in the periodic impulse response estimation for Example 1.}
\label{fig:1}
\end{figure}

\begin{figure*}[htbp]
\centering
\includegraphics[width=6.8in]{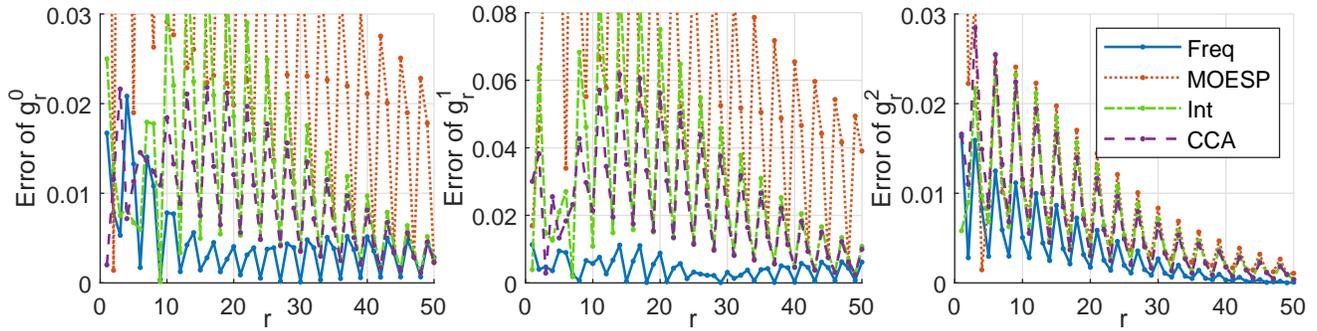}
\caption{Errors in the periodic impulse response estimation for Example 2.}
\label{fig:2}
\end{figure*}

The identification results are shown in Figures~\ref{fig:1} and \ref{fig:2} for Examples~1 and 2 respectively, in terms of the absolute estimation errors of the periodic impulse responses $g^\tau_r$, as the state-space matrices are only equivalent up to unknown similarity transforms. In Example 1, the system is autonomous at $\tau=1$, so only the impulse responses at $\tau=0$ are shown. As can be seen from both figures, the estimation error of the proposed method is smaller than the other three time-domain methods. In particular for Example 2, the time-domain methods fail to provide a meaningful estimation of the system, whereas the proposed frequency-domain method is still able to obtain reasonable results.

\begin{figure}[htbp]
\centering
  \begin{tabular}{ c @{\hspace{5pt}} c }
    \includegraphics[width=1.55in]{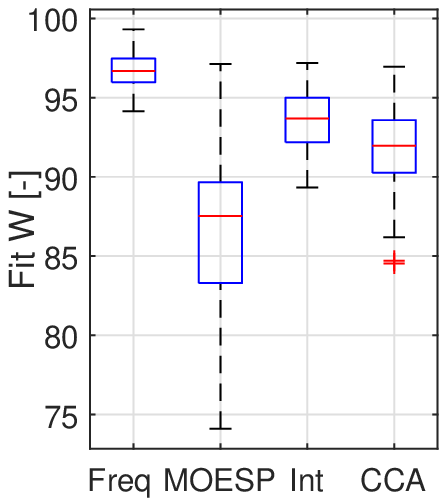} &
      \includegraphics[width=1.55in]{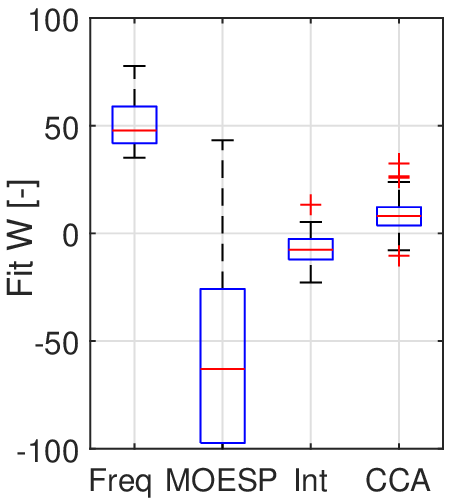} \\
    \footnotesize (a) Example 1&
      \footnotesize (b) Example 2
  \end{tabular}
\caption{Comparison of fitting performance with Monte Carlo simulations.}
\label{fig:3}
\end{figure}

To quantitatively assess the performance of the identification schemes, 100 Monte Carlo simulations with different noise realizations were conducted for both examples. The performances are parameterized by the following fitting metric
\begin{equation}
W=100\cdot \left(1-\left[\frac{\sum_{\tau=1}^P\sum_{r=1}^{n_g}(g^{\tau}_r-\hat{g}^{\tau}_r)^2}{\sum_{\tau=1}^P\sum_{r=1}^{n_g}(g^{\tau}_r-\bar{g})^2}\right]^{1/2}\right),
\end{equation}
where $g^{\tau}_r$ are the true impulse response coefficients, $\hat{g}^{\tau}_r$ are the estimated coefficients, $\bar{g}$ is the mean of true coefficients, and $n_g$ is selected as 50 here. 
The box plots of the metric $W$ for both examples are shown in Fig.~\ref{fig:3}. In both examples, the proposed method has a better fitting performance compared to the time-domain method.

The above results demonstrate that the proposed method performs better than the time-domain methods when periodic input-output data are available. This advantage is mainly due to the fact that it makes use of the periodic nature of the identification data. This gives the complete input history of the system or, in other words, the initial condition, whereas in the time-domain method, past inputs are assumed unknown.

\begin{figure}[htbp]
\centering
\includegraphics[width=3.2in]{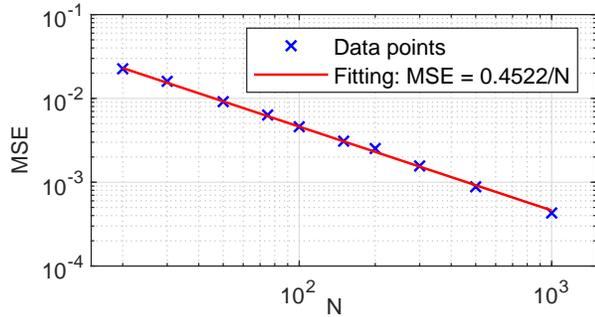}
\caption{Mean squared error of the frequency-domain subspace estimate under different data lengths.}
\label{fig:4}
\end{figure}

Finally, we demonstrate the consistency property that is proved in Theorem~\ref{thm:1} by conducting Monte Carlo simulations of Example 1 with increasing data length $N$. The results are shown in Fig.~\ref{fig:4} where the estimation error is characterized by the mean squared error of the periodic impulse response estimate. It can been seen that estimate is consistent with a convergence rate of $1/N$.

\section{Conclusions}
\label{sec:conc}

In this paper, we have proposed an LTP subspace identification method designed for periodic identification data. This method applies a two-step approach: first the generalized ETFE of the lifted LTP system is obtained from the identification data; then the time-aliased periodic impulse response derived from the lifted frequency response is used to construct an order-revealing decomposition of the original LTP system, from which the general framework of subspace identification can be utilized. The proposed algorithm complements the available subspace identification algorithms for LTP systems, and shows an advantage in model fitting from numerical simulation when periodic data are available.

\IEEEtriggeratref{4}
\bibliographystyle{IEEEtran}
\bibliography{IEEEabrv,refs}

\end{document}